\pdfoutput=1
\documentclass[review,3p,times,fleqn,11pt]{elsarticle}
\usepackage{graphicx}
\usepackage{amssymb}
\usepackage{natbib}
\bibpunct[, ]{(}{)}{,}{a}{}{,}%
\usepackage{lineno}
\usepackage{url}
\usepackage{color}
\usepackage{amsmath}
\usepackage{algorithm}  
\usepackage{algorithmic} 
\usepackage{appendix}
\usepackage{booktabs}
\usepackage{setspace}
\usepackage{subfigure}
\usepackage{tabularx}
\usepackage{lscape}
\usepackage{multirow}
\usepackage{tabularx}
\usepackage{makecell}
\usepackage{orcidlink}
\usepackage{ragged2e}
\usepackage{graphicx} 
\usepackage{float} 
\usepackage{booktabs}
\usepackage{multirow}
\usepackage{adjustbox}
\newcolumntype{Y}{>{\raggedright\arraybackslash}X} 
\newcommand{\mycomment}[1]{}

\journal{XXXX}

\begin{document}
	
	\begin{frontmatter}

		\title{New Adaptive Mechanism for Large Neighborhood Search using Dual Actor-Critic}

		\author[label1]{Shaohua Yu~\orcidlink{0000-0002-2110-4621}}
		\ead{shaohua.yu@njust.edu.cn}
            \author[label2]{Wenhao Mao}
            \ead{maowenhao32@gmail.com}
            \author[label2]{Zigao Wu*}
            \ead{zgwu@ncepu.edu.cn}
		\author[label3,label4]{Jakob Puchinger}
		\ead{jpuchinger@em-normandie.fr}

		\address[label1]{School of Intelligent Science and Technology, Nanjing University of Science and Technology, Nanjing 210094, China}
        \address[label2]{Department of Mechanical Engineering, North China Electric Power University, Baoding 071003, China}
  		\address[label3]{EM Normandie Business School, Métis Lab, 92110, Clichy, France}
		\address[label4]{Université Paris-Saclay, CentraleSupélec, Laboratoire Génie Industriel, 91190, Gif-sur-Yvette, France}

\begin{abstract}

Adaptive Large Neighborhood Search (ALNS) is a widely used heuristic method for solving combinatorial optimization problems. ALNS explores the solution space by iteratively using destroy and repair operators with probabilities, which are adjusted by an adaptive mechanism to find optimal solutions. However, the classic ALNS adaptive mechanism does not consider the interaction between destroy and repair operators when selecting them. To overcome this limitation, this study proposes a novel adaptive mechanism. This mechanism enhances the adaptability of the algorithm through a Dual Actor-Critic (DAC) model, which fully considers the fact that the quality of new solutions is jointly determined by the destroy and repair operators. It effectively utilizes the interaction between these operators during the weight adjustment process, greatly improving the adaptability of the ALNS algorithm. In this mechanism, the destroy and repair processes are modeled as independent Markov Decision Processes to guide the selection of operators more accurately. Furthermore, we use Graph Neural Networks to extract key features from problem instances and perform effective aggregation and normalization to enhance the algorithm’s transferability to different sizes and characteristics of problems. Through a series of experiments, we demonstrate that the proposed DAC-ALNS algorithm significantly improves solution efficiency and exhibits excellent transferability.

\end{abstract}
    \begin{keyword}
 Adaptive large neighborhood search; Deep reinforcement learning; Actor-Critic model; Graph neural network
    \end{keyword}
\end{frontmatter}

\section{Introduction}

Metaheuristic algorithms are a type of algorithm used to solve optimization problems, especially those that are NP-hard. They don't guarantee finding the optimal solution, but they usually find a near-optimal solution in a reasonable amount of time. Because of their efficiency and flexibility in handling complex problems, metaheuristics have become popular for solving real-world problems. The Adaptive Large Neighborhood Search (ALNS) algorithm is a widely used metaheuristic algorithm proposed by \cite{ropke2006adaptive}, and has been extensively applied in various optimization problems in logistics, manufacturing, and other fields.

The core idea of the ALNS algorithm is to build new solutions by alternately using different destroy and repair operators. Destroy operators are responsible for removing parts of the current solution, while repair operators focus on reconstructing the remaining parts. The selection probability of these operators is adaptively adjusted through a weight adjustment mechanism and is selected using an algorithm similar to a roulette wheel. ALNS combines the intuitiveness of neighborhood search with the flexibility of adaptive mechanisms, allowing it to adjust the search strategy according to problem characteristics automatically. However, numerous numerical experiments have revealed an important phenomenon: in the classical ALNS algorithm, the effect of the adaptive mechanism is less significant than expected. According to  \cite{turkevs2021meta}, a classical adaptive mechanism into the ALNS algorithm only increases the objective function value by an average of 0.14\%. This finding suggests that although classical adaptive mechanisms have theoretical advantages, their practical improvement is limited. Therefore, designing more efficient adaptive mechanisms is crucial.

In the classical adaptive mechanism, selecting destroy or repair operators does not fully consider the interaction between these two types of operators. The selection process mainly relies on the operators' immediate weight within their respective destroy or repair sets. For example, when selecting a repair operator, the type of the previously used destroy operator and its impact on the problem solution are not considered. This independent selection strategy may lead to improper combinations of operators, affecting the algorithm's performance. Additionally, the classical adaptive mechanism relies too heavily on a single indicator: the operator's weight within the set and its immediate contribution to the objective function. This approach neglects the potential impact of the current state on the problem solution throughout the solving process, which may not fully assess the actual effect of the operators and thus affect the overall efficiency of the adaptive mechanism.

To overcome these limitations, a more detailed and comprehensive adaptive strategy is needed. This strategy should consider the mutual dependence between destroy and repair operators and the impact of the overall solution process's state on operator selection. Therefore, this paper proposes a new adaptive mechanism for the ALNS algorithm based on the Dual Actor-Critic Model (DAC), which can better perform the adaptive selection of operators. The new adaptive mechanism constructs policy networks for the destroy and repair processes, using value networks to evaluate the choice of operators, thus formulating better optimization strategies. In this process, we use graph neural networks (GNNs) to extract path features as the state space for the destroy process and consider the selected destroy operator and the features extracted by the GNNs as the state space for the repair process to evaluate better the potential impact of the current state on operator selection during the solution process.

The main contributions of this paper are as follows:

\begin{enumerate}
    \item We use DAC to optimize the adaptive mechanism of ALNS, effectively use the mutual influence between destroy and repair operators during the operator selection process, and fully consider the current state's potential impact on problem-solving.
    \item We use GNNs to extract and aggregate graph features, constructing the state spaces for both the destroy and repair Actor networks, allowing the network to adapt to problem instances of different sizes and characteristics, thereby improving the adaptability and transferability of the algorithm.
    \item We conduct extensive numerical experiments on the proposed adaptive mechanism using the standard datasets for the Capacitated Vehicle Routing Problem (CVRP) and the Vehicle Routing Problem with Time Windows (VRPTW), verifying the effectiveness of the proposed method.
\end{enumerate}

The remaining article structure follows: Section 2 introduces the related work, Section 3 is the methodology, Section 4 is numerical experiments, and Section 5 conclude.

\section{Related Work}

\subsection{Adaptive Large Neighborhood Search}

The ALNS algorithm is an efficient metaheuristic approach that enhances the large neighborhood search algorithm \citep{shaw1998using} with adaptive mechanisms. It uses a weight adjustment mechanism and a roulette wheel selection method to choose different operators adaptively, guiding the destroy and repair processes of feasible solutions, thus achieving better global optimization capabilities \citep{ropke2006adaptive,Lutz2015AdaptiveLN,mourad2021integrating}.

Since the proposal of the ALNS algorithm, scholars have sought various methods to optimize its standard components, including initial solution generation, destroy and repair operators, adaptive mechanisms, and acceptance and termination criteria.

In ALNS, the generation of initial solutions and the selection of destroy and repair operators often exhibit strong problem-specific characteristics, and their design must be closely dependent on the structural features of the problem’s solutions. For the design of initial solution generation methods, most studies use a simple heuristic approach to obtain an initial solution. Some scholars have proposed multi-start ALNS methods, generating multiple initial solutions simultaneously to increase the diversity of initial solutions and thus enhance the diversity of search directions \citep{han2016multi,marti2019intelligent}. Typical ones for the design of destroy and repair operators include random removal, greedy removal, random insertion, and greedy insertion \citep{qu2013heterogeneous}. Some studies select destroy and repair operators independently through a roulette wheel mechanism, such as in the standard form of ALNS \citep{ropke2006adaptive}. Others have designed pairs of destroy-repair operators, enabling their combined use in ALNS \citep{belhaiza2019hybrid}. According to \cite{yu2025hybrid}, the former outperforms the latter.

The adaptive mechanisms, acceptance, and termination criteria in ALNS are usually general-purpose. According to \cite{mara2022survey}, there are many studies on acceptance and termination criteria in ALNS. For example, acceptance criteria based on Greedy Mechanism, Metropolis Criterion, Record-to-Record, Threshold Acceptance, and Pareto Dominance \citep{santini2018comparison}, as well as termination criteria based on Number of Iterations, Number of Non-Improving Iterations, Running Time Limit, and Annealing Temperature \citep{breunig2019electric,yu2022van}, are widely used in ALNS. However, current research on the adaptive mechanisms in ALNS is limited. Out of 251 relevant articles surveyed in \cite{mara2022survey}, 250 used the roulette wheel as the adaptive mechanism, accounting for 99.21\%, and only one used the stochastic universal sampling strategy \citep{chowdhury2019modified}, accounting for 0.40\%. In the past two years, research on adaptive mechanisms has also been limited, with \cite{yu2025hybrid} proposing an ALNS algorithm adaptive mechanism based on the destroy-based repair weight-adjusted method.

Although research indicates that the types of adaptive mechanisms currently studied are relatively few, with the weight-adjusted roulette wheel being the primary adaptive mechanism, this does not mean that existing adaptive mechanisms perform exceptionally well. A significant phenomenon revealed by numerous numerical experiments is that the effectiveness of adaptive mechanisms in classical ALNS algorithms has not met expectations. According to \cite{turkevs2021meta}, the classical adaptive mechanism into the ALNS algorithm only results in an average improvement of 0.14\% in the objective function value. This finding suggests that although adaptive mechanisms have theoretical advantages, their practical effectiveness is limited. Therefore, designing more efficient adaptive mechanisms is particularly important.

\subsection{Deep reinforcement learning in ALNS}

Artificial intelligence methods are innovative optimization techniques that simulate and understand human intelligence, behavior, and laws by constructing computer systems. \cite{aydin2024feature} indicate that for metaheuristic algorithms, problems can be represented by multiple features determined through landscape analysis, and artificial intelligence technologies can be used to build adaptive operator selection schemes to handle efficiency and scalability problems.

Researchers have recently explored applying reinforcement learning techniques to the operator selection process in metaheuristic algorithms. \cite{kalatzantonakis2023reinforcement} demonstrates that reinforcement learning can optimize the parameter settings of ALNS, including the weights of neighborhood structures and the shaking strategy, by learning the environment and reward mechanisms, thereby enhancing the efficiency and quality of the search. \cite{durgut2021adaptive} proposes a Q-learning-based reinforcement learning algorithm integrated into the standard artificial bee colony algorithm to guide the algorithm’s operation selection module, solving the set union knapsack problem. The core of this method is to map problem states to a set of best-matching operators, increasing diversity throughout the search process and constructing an optimal operator sequence. Experimental results validate the effectiveness of this approach.

Deep learning excels at extracting useful features from raw, high-dimensional data, enhancing state representation capabilities, and allowing algorithms to handle more complex problems and better adapt to different environments \citep{mnih2015human}. \cite{kallestad2023general} introduces deep learning methods based on reinforcement learning and proposes a deep reinforcement learning-assisted ALNS algorithm. This method uses the proximal policy optimization algorithm \citep{schulman2017proximal} to learn the strategy of selecting destroy-repair operator pairs. 
Experiments show the effectiveness of the deep reinforcement learning-based adaptive method on problems, including the CVRP. \cite{reijnen2024online} further uses deep reinforcement learning to select destroy-repair operator pairs, adjust parameters, and control acceptance criteria during the search process. The proposed method aims to learn from the search state and configure ALNS for the next iteration to obtain more effective solutions. They evaluated the method on a direction problem with random weights and time windows (proposed in the IJCAI competition). The results show that their method outperforms the classical ALNS.

Methods based on integrating deep reinforcement learning and graph neural networks have also received attention for optimizing ALNS. \cite{johnn2023graph} emphasizes the potential of graph neural networks in generalizing to larger problem instances than those observed during training, indicating that this method performs better than the classical ALNS adaptive layer.

Although the Q-learning algorithm is simple in principle, accessible to implement, and theoretically converges to the optimal policy, it may face the curse of dimensionality for problems with large-scale state spaces, meaning the Q-table becomes extremely large, making computation and maintenance difficult. The deep Q-network algorithm addresses this by introducing a deep neural network to approximate the Q-value function, which is suitable for significant state space problems. Still, it exhibits overestimation during training, reducing the learning efficiency of the agent.

We also note that existing ALNS algorithms integrated with deep reinforcement learning model the Markov Decision Process only for the selection of destroy-repair operator pairs \citep{kallestad2023general,reijnen2024online} or only for the destroy operators \citep{johnn2023graph}, and then use deep reinforcement learning algorithms for selection. However, according to \cite{yu2025hybrid}, classical ALNS adaptive mechanisms are superior to the destroy-repair operator pair approach but inferior to the destroy-based repair weight-adjusted approach, which considers the interaction between destroy and repair operators when making selections.

Therefore, fully considering the joint impact of destroy and repair operators on the quality of new solutions, modeling the destroy and repair processes separately as Markov Decision Processes, and giving full play to the role of destroy and repair operators in the weight adjustment process of the ALNS algorithm, could significantly benefit the solution of ALNS.

Proximal Policy Optimization (PPO) and Actor-Critic Model (AC) improve training stability and accelerate the convergence to a better policy by reducing variance during training. Notably, the AC algorithm can be seen as a special form of PPO \citep{huang2022a2c} in some contexts. The structure of the AC algorithm is highly flexible, consisting of two mutually independent parts: the actor (responsible for learning the policy) and the critic (responsible for evaluating the value function). This structural design allows us to customize and optimize the two networks separately to better adapt to specific needs. Therefore, designing a new adaptive mechanism based on DAC holds great potential.

\section{Methodology}

In solving VRP or other optimization problems, ALNS is an effective metaheuristic method, but its performance heavily relies on selecting destroy and repair operators. The adaptive mechanism of traditional ALNS is only based on the immediate weight of each operator, ignoring the influence of the current state on operator selection and the problem's solution, which may affect operator evaluation and algorithm efficiency. To address this, we propose an adaptive mechanism based on Dual Actor-Critic to improve the ALNS algorithm. This mechanism combines the graph neural network to construct policy networks for the destroy and repair processes separately and uses the value network to evaluate the operator selection process. The mechanism uses the path features extracted by the graph neural network as the state space of the destroy process, and the selected destroy operator and the features extracted by the graph neural network together constitute the state space of the repair process. The DAC-ALNS algorithm will more comprehensively consider the mutual dependence between the selection of destroy and repair operators, including the potential impact of the selection of destroy operators on the subsequent selection of repair operators, and better use the data in the entire solution process to guide operator selection.

In this section, we will first introduce the basic architecture of the DAC-ALNS algorithm, and secondly, we will introduce the Markov decision process involved in the DAC-ALNS algorithm.

\subsection{DAC-ALNS algorithm architecture}

In this section, we first introduce the Actor-Critic algorithm and how to select operators through DAC, and then introduce the algorithm framework of DAC-ALNS.

\bigskip
\textbf{(1) Actor-Critic Algorithm}

In the Actor-Critic algorithm, the Actor network’s role is to learn policies. It establishes a state space through the interaction between the agent and the environment and selects actions based on the policy. On the other hand, the Critic network is responsible for evaluating the policies learned by the Actor, guiding the Actor network to improve and optimize the policies continuously. The Critic network guides the policy updates by estimating the state-action value function. Therefore, the roles of Actor and Critic are complementary, interacting with and dependent on each other.

\begin{figure}[htbp]
    \centering
    \begin{minipage}[t]{0.48\linewidth}
        \centering
        \includegraphics[height=5cm]{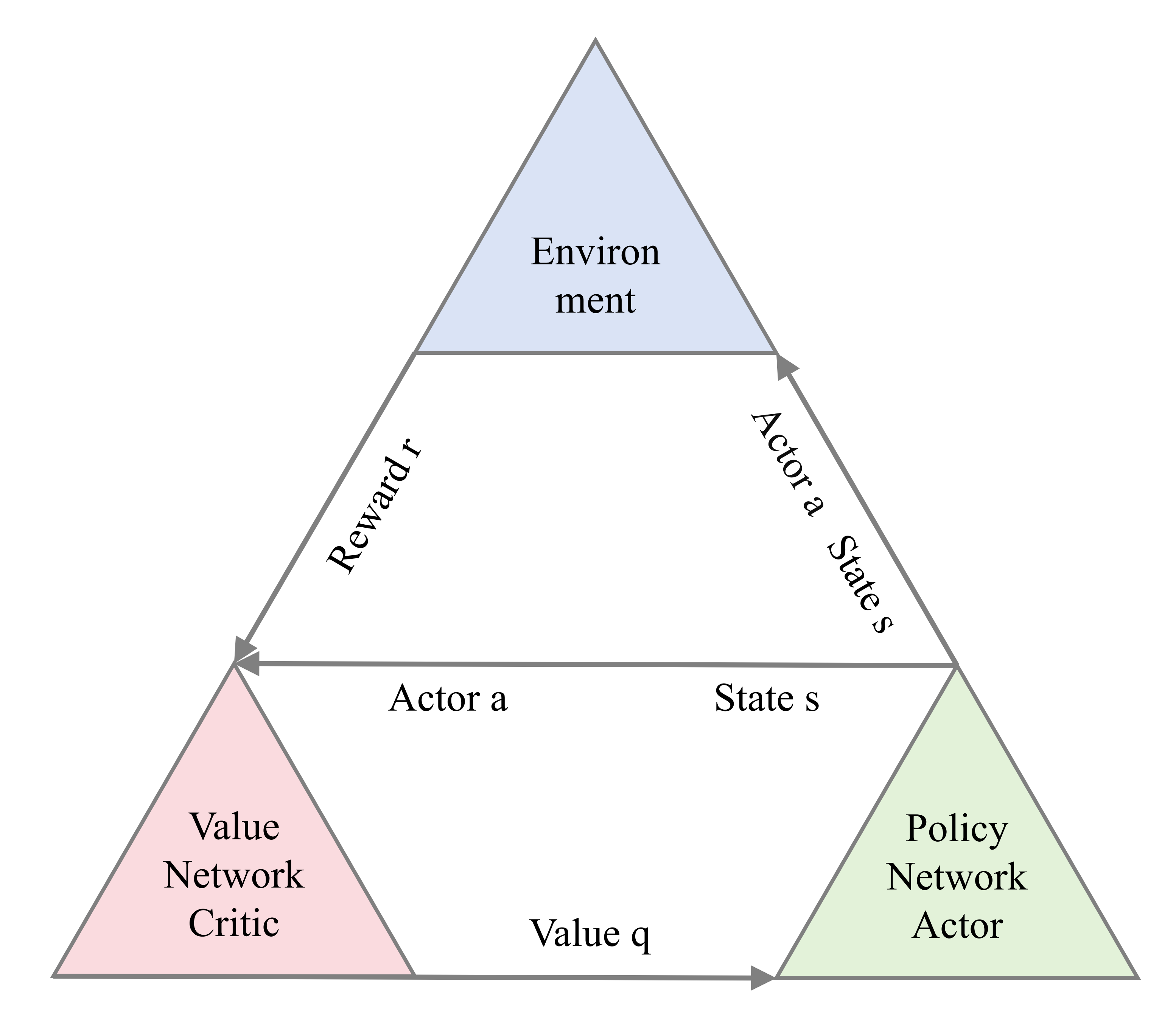} 
        \caption{Actor-Critic algorithm diagram}
        \label{fig:enter-label1}
    \end{minipage}
    \hfill
    \begin{minipage}[t]{0.48\linewidth}
        \centering
        \includegraphics[height=5cm]{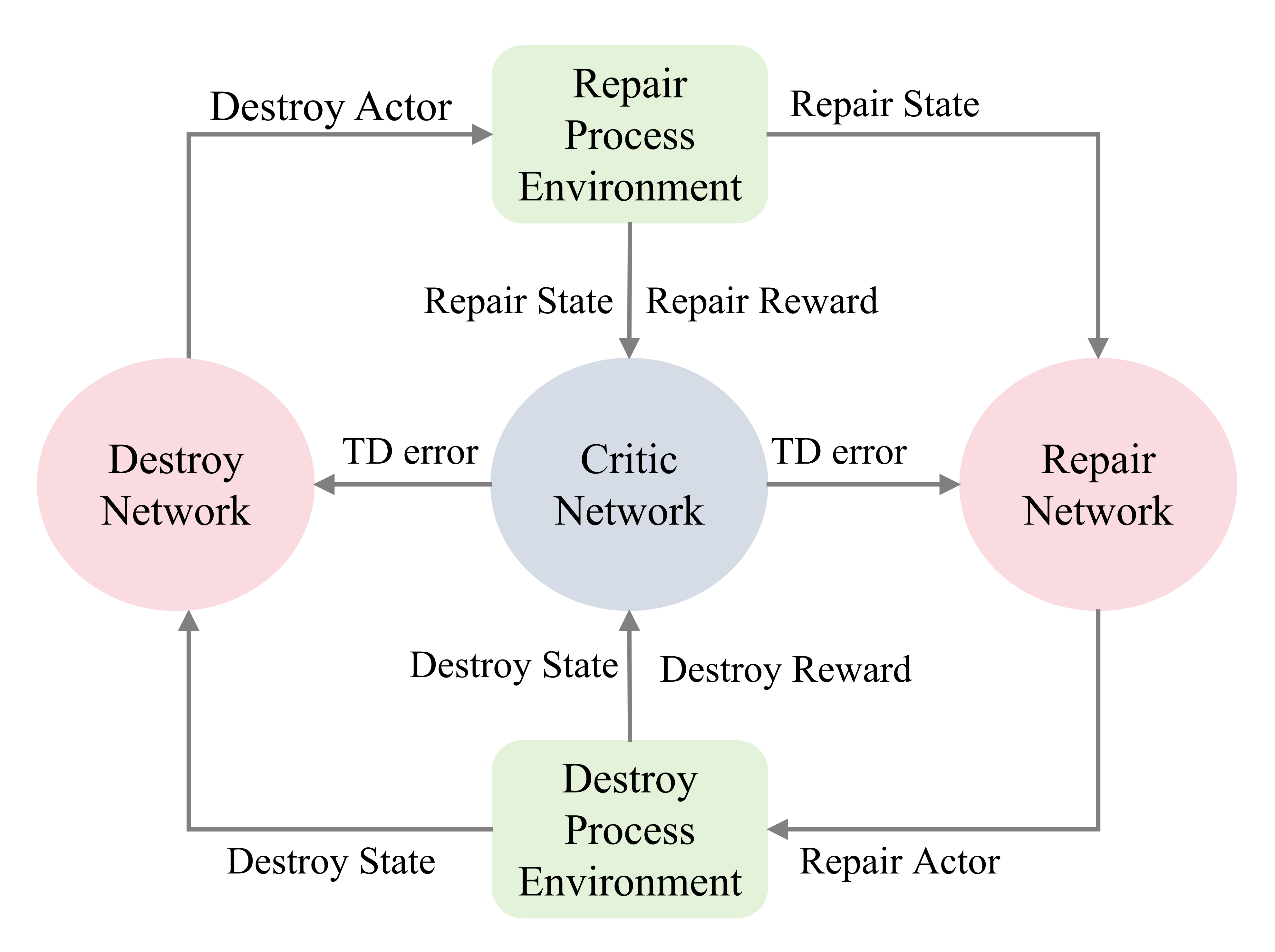}
        \caption{Operator weight adjustment mechanism based on DAC}
        \label{fig:enter-label2}
    \end{minipage}
\end{figure}

Figure \ref{fig:enter-label1} shows a schematic of the Actor-Critic algorithm.

\bigskip
\textbf{(2) DAC-based ALNS operator selection}

To enhance the effectiveness of the adaptive mechanism, we leverage the high scalability of the Actor-Critic algorithm components, modeling the destroy and repair operators as two separate neural networks. Specifically, we designed an Actor network for both the destroy and repair processes, each responsible for selecting the most suitable strategy. Additionally, these two Actor networks share a Critic network, which evaluates the selection of these strategies and assesses the current state. Consequently, we adopted the Dual Actor-Critic approach to replace the adaptive layer in the classic ALNS algorithm, more comprehensively considering the interdependence between the destroy and repair operators and more effectively using the data from the entire solution process to influence operator selection.

The mechanism for adjusting operator weights based on DAC is shown in Figure \ref{fig:enter-label2}.

\bigskip
\textbf{(3) DAC-ALNS algorithm framework}

We embed the DAC-based operator adaptive selection mechanism into the adaptive layer of ALNS (The ALNS algorithm architecture draws on the experience from \cite{Lutz2015AdaptiveLN}), resulting in the Dual Actor-Critic-based adaptive large neighborhood search algorithm. 

The pseudo-code for the training and deployment phases of the proposed DAC-ALNS are provided in Algorithm \ref{A:Training-DAC-ALNS} and Algorithm \ref{A:Deployment-DAC-ALNS}, respectively.

\begin{algorithm}[htbp] 
	\caption{Training Phase: Dual Actor-Critic ALNS}
    \label{A:Training-DAC-ALNS}
	\begin{algorithmic}[1]  
        \STATE \textbf{Training}
		\STATE \textbf{Input:} Number of training steps $M$
		\STATE $Step \leftarrow 0$
		\WHILE{$Step < M$}
		\STATE Initialize problem instance
		\STATE Initialize feasible solution $x$
		\STATE $x_{best} \leftarrow x$
		\STATE Initialize state $s_{t_1}$ $\leftarrow$ 0, $s_{t_2}$ $\leftarrow$ 0
		\STATE Initialize Actor network parameters $\theta$ and Critic network parameters $\omega$
		\WHILE{not satisfying the stopping condition}
		\STATE Select and execute the destroy operator $d$ with policy $\pi_{\theta_1}$ based on state $s_{t_1}$
		\STATE Update state $s_{t_2+1}$ with the destroyed solution and node, and obtain reward $r_{t_1}$
		\STATE Select the repair operator $r$ with policy $\pi_{\theta_2}$ based on state $s_{t_2+1}$
		\STATE $x_t \leftarrow r(d(x))$
		\IF{accept ($x_t, x$)}
		\STATE $x \leftarrow$ $x_t$
		\ENDIF
		\IF{c($x_t$) $<$ c($x_{best}$)}
		\STATE $x_{best}$ $\leftarrow$ $x_t$
		\ENDIF
		\STATE $Step$ $\leftarrow$ $Step$ + 1
		\STATE Update state $s_{t_1+1}$ and obtain reward $r_{t_2}$
		\STATE Update Critic network parameters $\omega$ using TD
		\STATE Update Actor network parameters $\theta$ using gradient ascent
		\ENDWHILE
		\ENDWHILE
        \RETURN Optimized parameters $\theta_1, \theta_2$
    \end{algorithmic}  
\end{algorithm}
\begin{algorithm}[htbp] 
	\caption{Training Phase: Dual Actor-Critic ALNS}
    \label{A:Deployment-DAC-ALNS}
    \begin{algorithmic}[1]  
		\STATE \textbf{Deployment:}
		\STATE \textbf{Input:} Selection policy $\pi_\theta$
		\STATE Feasible solution $x$; $x_{best} \leftarrow x$
		\WHILE{not satisfying the stopping condition}
		\STATE Select the destroy operator $d \in$ $\Omega^-$ with policy $\pi_{\theta_1}$ based on state $s_{t_1}$
		\STATE Select the repair operator $r \in$ $\Omega^+$ with policy $\pi_{\theta_2}$ based on state $s_{t_2}$
		\STATE $x_t \leftarrow r(d(x))$
		\IF{accept ($x_t, x$)}
		\STATE $x \leftarrow x_t$
		\ENDIF
		\IF{$c(x_t) < c(x_{best}$)}
		\STATE $x_{best} \leftarrow x_t$
		\ENDIF
		\ENDWHILE
		\STATE \textbf{return:} $x_{best}$
	\end{algorithmic}  
\end{algorithm}

\subsection{Markov Decision Processes}

The mathematical foundation of reinforcement learning is Markov Decision Processes (MDPs). An MDP typically consists of a state space, an action space, and a reward function, among other components. Reinforcement learning is a sequential decision-making process that seeks to find a decision rule (i.e., policy) that maximizes the cumulative reward of the system, or in other words, maximizes the value \citep{nazari2018reinforcement}.

Next, we will introduce the state space, action space, and reward function involved in the DAC-ALNS algorithm for CVRP and VRPTW.

\subsubsection{State Space}

In the Markov Decision Process, the state representation encapsulates environmental information and dynamic changes perceived by the agent, providing essential input for decision-making and reward assessment within deep reinforcement learning algorithms. The design quality of state space directly influences both convergence speed and the ultimate performance of the learning algorithms.

In this paper we employ GNNs to extract compact, informative graph-level features, significantly reducing state complexity while retaining key characteristics.

\paragraph{State Representation (CVRP)}

We represent each CVRP solution (partial or complete) as a graph $G=(V,E)$, with node features comprising planar coordinates, demand, and a binary indicator of service completion. The edge set integrates a static $k$-NN or fully connected topology with dynamically induced arcs, using inverse distance for edge weighting. Node features pass through a two-layer Graph Convolutional Network (GCN) to produce embeddings, which are subsequently aggregated via attention-weighted average pooling and element-wise max pooling into a concise graph-level embedding $g\in\mathbb{R}^{2d}$. 

To incorporate search-phase context, we append two scalars: the destruction ratio $\rho$ (fraction of customers removed) and the normalized identifier $\tilde d$ of the destroy operator. Thus, the destroy-state (before removal) is defined as $s^{\mathrm{des}}=[g|0|0]$, and the repair-state (after removal) as $s^{\mathrm{rep}}=[g'|\rho|\tilde d]$, with $g'$ recalculated from the partially modified solution. The resultant embedding supports both the shared critic and actor networks.

\paragraph{State Representation (VRPTW)}

Similar to the CVRP, each VRPTW solution (partial or complete) is encoded as a graph $G=(V,E)$, integrating a static $k$-NN (or fully connected) backbone with dynamic route-derived arcs. Edges use either inverse or direct distance weighting. Nodes carry 10-dimensional features combining spatial coordinates, demand, service status, normalized time window constraints, service durations, and temporal metrics such as tightness and waiting times. 

A two-layer GCN processes these features into embeddings, which are aggregated via attention-weighted average and element-wise max pooling, yielding a graph-level embedding $g\in\mathbb{R}^{2d}$. To capture solution modification context, we again include the destruction ratio $\rho$ and the normalized destroy operator id $\tilde d$, defining the destroy-state as $s^{\mathrm{des}}=[g|0|0]$ and the repair-state as $s^{\mathrm{rep}}=[g'|\rho|\tilde d]$, where $g'$ reflects the partially destroyed solution state.

\bigskip
The state-space summary under the DAC-ALNS is shown in Table \ref{tab:state-space-transposed}.

\begin{table}[htbp]
\centering
\small 
\setlength{\tabcolsep}{6pt}
\renewcommand\arraystretch{1.4}
\caption{State-space summary under the DAC-ALNS framework}
\label{tab:state-space-transposed}
\begin{tabularx}{\linewidth}{l >{\hsize=0.85\hsize}X >{\hsize=1.15\hsize}X}
\toprule
\textbf{Aspect} & \textbf{CVRP} & \textbf{VRPTW} \\
\midrule

\textbf{Graph \& Edges} & 
$G=(V,E)$; static $k$-NN + route-induced dynamic arcs; edge weight $w_{uv} = 1/d_{uv}$. & 
$G=(V,E)$; static $k$-NN + dynamic arcs from current routes; edge weight $w_{uv} = 1/d_{uv}$. \\

\addlinespace[5pt]
\textbf{Node Features} & 
$[x_i, y_i, q_i, \delta_i]$ \newline 
\textit{Dim:} 4 \newline
\textit{Note:} $\delta_i \in \{0,1\}$ (served status). & 
$[x_i, y_i, q_i, \delta_i, \frac{e_i-S}{H}, \frac{l_i-S}{H}, \frac{w_i}{\|w\|_\infty}, \frac{s_i}{\|s\|_\infty}, \frac{t_i}{H}, \frac{wt_i}{H}]$ \newline 
\textit{Dim:} 10 \newline
\textit{Note:} $H=E-S$ (time horizon). \\

\midrule
\multicolumn{3}{c}{\textit{Common Components for Both Frameworks}} \\
\midrule

\textbf{Embedding} & 
\multicolumn{2}{p{0.85\linewidth}}{Two-layer GCN; attention-weighted average and element-wise max pooling $\Rightarrow g \in \mathbb{R}^{2d}$.} \\

\textbf{Context Scalars} & 
\multicolumn{2}{p{0.85\linewidth}}{Destruction ratio $\rho \in (0,1)$; normalized destroy operator index $\tilde{d}$.} \\

\textbf{State Vectors} & 
\multicolumn{2}{p{0.85\linewidth}}{
    $s^{\mathrm{des}} = [g \parallel 0 \parallel 0]$ (initial state); \newline
    $s^{\mathrm{rep}} = [g' \parallel \rho \parallel \tilde{d}]$ (intermediate state, $g'$ from partial solution).
} \\
\bottomrule
\end{tabularx}
\end{table}

\subsubsection{Action Space}

The action space of the DAC-ALNS algorithm consists of two fundamental actions: destroy actions and repair actions. Destroy actions involve removing a portion of elements from the current solution, while repair actions add elements to the solution after destruction, forming a new feasible solution. These two actions together define the action space of deep reinforcement learning, which is the set of actions the algorithm can choose to take at each step. In addition, we introduce a destroy scale strategy that enables the DAC-ALNS algorithm to adjust the destroy scale.

\bigskip

Our destroy and repair actions are adapted to problem-specific constraints. Since the operator sets for CVRP and VRPTW are identical, they are summarized as follows: Random removal, worst removal, random repair, greedy repair, regret repair. For details, please see \ref{A:dac}.

\subsubsection{Reward Function}

We use the total route cost as the core signal (lower is better). Let
$C_{\mathrm{before}}$ denote the cost before applying the current operator,
$C_{\mathrm{destroy}}$ the cost immediately after \emph{destroy},
$C_{\mathrm{repair}}$ the cost immediately after \emph{repair},
and $C_{\mathrm{best}}$ the current global best (best-so-far) cost.

\bigskip
\textbf{(1) Destroy Reward Function}

The destroy actor removes part of the solution to create room for subsequent repair. Its reward combines a quality flag and a potential term. 
In our implementation, we define two base reward parameters $\alpha_1$ and $\alpha_2$. If the solution's total cost after destruction is lower than or equal to the current global best (indicating high potential), a higher base reward $\alpha_2$ is given. If the cost is lower than the current solution but has not reached the global best, a base reward $\alpha_1$ is applied. Additionally, a potential term scaled by $\beta$ is added to reflect the gap from the global optimum. If the cost increases after destruction, the reward is zero.
The reward function for the destroy process is defined as follows:

\begin{equation}
R_{\mathrm{destroy}}=
\begin{cases}
\alpha_{1}
+\beta\,\dfrac{C_{\mathrm{best}}-C_{\mathrm{destroy}}}{\max(\varepsilon,\,C_{\mathrm{best}})},
& \text{if } C_{\mathrm{best}}<C_{\mathrm{destroy}}<C_{\mathrm{before}},\\[8pt]
\alpha_{2}
+\beta\,\dfrac{C_{\mathrm{best}}-C_{\mathrm{destroy}}}{\max(\varepsilon,\,C_{\mathrm{best}})},
& \text{if } C_{\mathrm{destroy}}\le C_{\mathrm{best}},\\[8pt]
0, & \text{if } C_{\mathrm{destroy}} \ge C_{\mathrm{before}}
\end{cases}
\end{equation}

\bigskip
\textbf{(2) Repair Reward Function}

The repair actor reconstructs the solution, and we combine a quality-improvement term with an exploration bonus. The exploration component promotes less frequently used repair operators via
\begin{equation}
\mathrm{explore}
=\delta\!\left(1-\frac{f_{\mathrm{rep}(id)}}{\max\!\bigl(1,\,f_{\max}\bigr)}\right),
\end{equation}
where $\delta\ge0$ controls the exploration strength, $f_{\mathrm{rep}(id)}$ is the usage count of the current repair operator with index \texttt{id}, and $f_{\max}=\max_j f_{\mathrm{rep}(j)}$ is the maximum usage count across all repair operators. For quality, we use three tiers with base values $r_1,r_2,r_3\in\mathbb{R}$ (typically $r_1 > r_2 > r_3$):

\begin{equation}
R_{\mathrm{repair}}=
\begin{cases}
r_{1}+\mathrm{explore},
& \text{if } C_{\mathrm{repair}}<C_{\mathrm{best}},\\[4pt]
r_{2}+\mathrm{explore},
& \text{if } C_{\mathrm{best}}\le C_{\mathrm{repair}}<C_{\mathrm{before}},\\[4pt]
r_{3}+\mathrm{explore},
& \text{if } C_{\mathrm{repair}}\ge C_{\mathrm{before}}.
\end{cases}
\end{equation}

In words, repair actors that beat the current best receive the largest positive base reward $r_1$; repair actors that improve on the pre-operation cost but do not surpass the best receive a smaller positive base reward $r_2$; and repair actors that fail to improve on the pre-operation cost receive a penalty or small reward $r_3$. The exploration bonus is then added in all three cases.

\section{Numerical Experiments}

To evaluate the DAC-ALNS model, we conducted a two-phase experimental study.First, we compared the algorithm against classic benchmarks to demonstrate the high quality of its solutions.Then, we conducted generalization experiments to confirm that the model maintains stability on unseen data and demonstrates scalability as problem sizes increase.

The frameworks of the proposed DAC-ALNS, Lutz’s (2015) improved classic ALNS, and the AC-ALNS baseline are depicted in Figure 3.

\begin{figure}[htbp]
    \centering
    \includegraphics[width=0.8\linewidth]{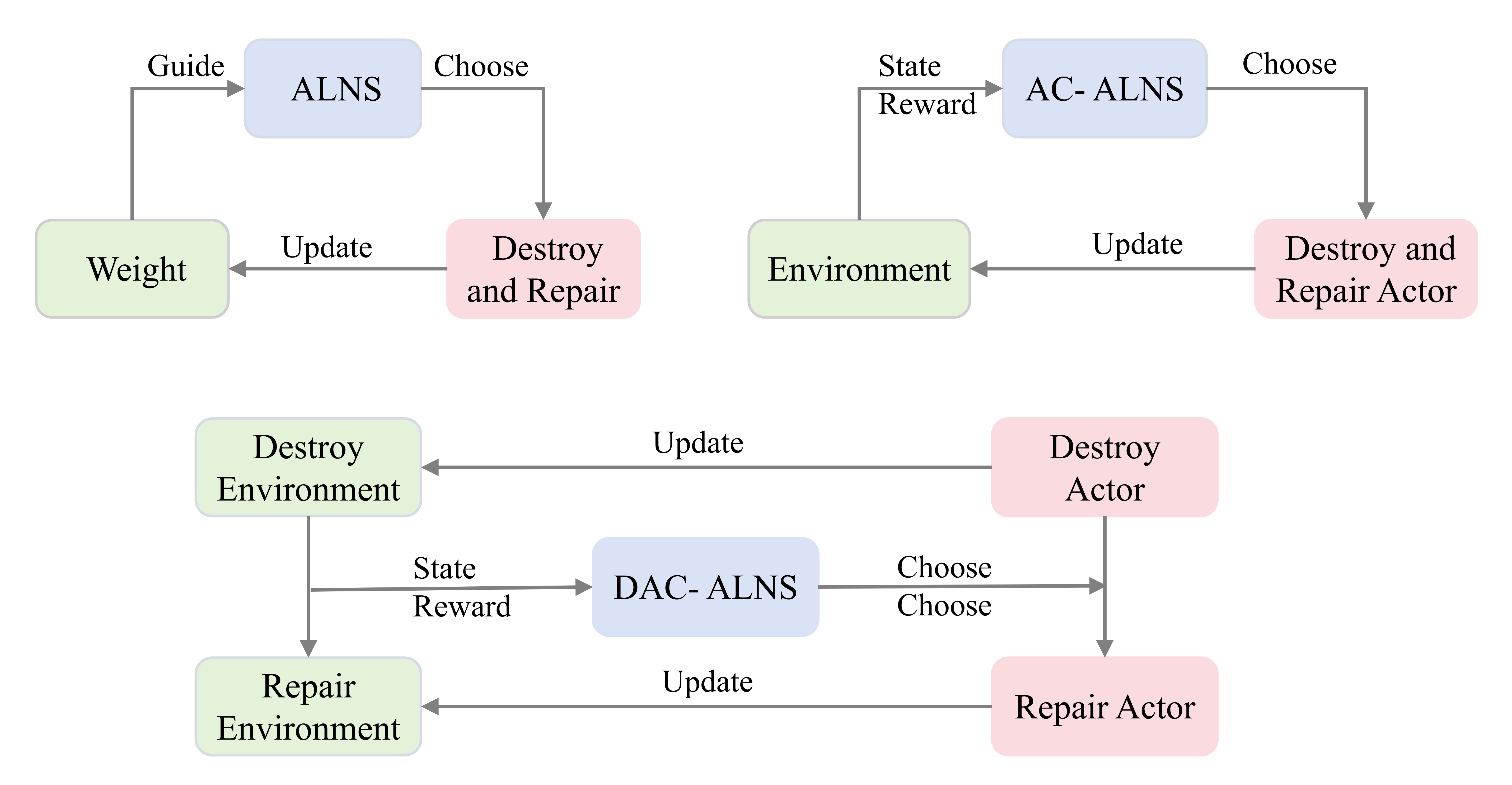}
    \caption{Schematic diagram of the operational logic for ALNS, AC-ALNS, and DAC-ALNS.}
    \label{fig:dac_com1}
\end{figure}

\subsection{Experimental Setup}

\subsubsection{Datasets and Benchmarks}

We adopted the classic CVRP and VRPTW datasets, which not only ensure experimental reproducibility but also allow subsequent researchers to compare their proposed algorithms with ours.

\begin{enumerate}
    \item \textbf{CVRP}: We utilized the classic instances proposed by \cite{augerat1995approche} (Sets A, B, and P), as their diverse customer distributions and capacity constraints help us fully evaluate the performance of DAC-ALNS.
    
    \item \textbf{VRPTW}: We opted for the renowned Solomon benchmark suite (Sets C, R, and RC) \citep{solomon1987algorithms}, which features Clustered, Random, and Mixed geographical distributions as well as diverse time windows. This choice allows for a more effective assessment of DAC-ALNS on VRPTW problems.
\end{enumerate}

\paragraph{Instance Stratification}
To provide a comprehensive assessment of DAC-ALNS across different problem scales, we categorize the test instances into three size-based categories---Small, Medium, and Large---determined by the number of customer nodes ($n$), as detailed in Table~\ref{tab:dataset-classification}.

\begin{table}[htbp]
\centering
\caption{Dataset split for  comparative experiments}
\label{tab:dataset-classification}
\begin{tabular}{lccc}
\toprule
\multirow{2}{*}{\textbf{Scale}} & \multirow{2}{*}{\textbf{Node Range ($n$)}} & \multicolumn{2}{c}{\textbf{No. of Instances}} \\
\cmidrule(lr){3-4}
 & & \textbf{CVRP} & \textbf{VRPTW} \\
\midrule
Small  & $n < 40$      & 18 & 18 \\
Medium & $40 \le n < 60$ & 18 & 18 \\
Large  & $60 \le n < 80$ & 18 & 18 \\
\bottomrule
\end{tabular}
\end{table}

Table~\ref{tab:generalization-data} summarizes the dataset split for the generalization experiments.

\begin{table}[htbp]
\centering
\caption{Dataset split for generalization analysis}
\label{tab:generalization-data}
\begin{tabular}{ll l}
\toprule
\textbf{Gen. Type} & \textbf{Problem} & \textbf{Transfer Protocol (Source $\to$ Target)} \\
\midrule
\multirow{2}{*}{\textbf{In-distribution}} 
 & CVRP  & Set A/B/P (Medium) $\to$ Remaining Set A/B/P \\
 & VRPTW & Sets C/R/RC (Medium) $\to$ Remaining Sets C/R/RC \\
\addlinespace[4pt] 
\multirow{2}{*}{\textbf{Out-of-distribution}} 
 & CVRP  & Set A (Small/Medium) $\to$ Set A (Large) \\
 & VRPTW & Set R (Small/Medium) $\to$ Set R (Large) \\
\bottomrule
\end{tabular}
\end{table}

Note for VRPTW, test instances were synthesized via random sampling from the original Solomon benchmarks to strictly adhere to the stratified node ranges ($n < 40$, $40 \le n < 60$, and $60 \le n < 80$). Furthermore, the generalization analysis utilizes datasets derived directly from the comparative set to maintain experimental coherence.

\subsubsection{Baselines}

To systematically assess DAC-ALNS, we conduct comparisons with two baselines, aiming to validate its solution quality for CVRP and VRPTW and to showcase the advantages of the Dual architecture over the  AC-ALNS architecture:
\begin{enumerate}
    \item ALNS (Baseline): A classic paradigm of traditional metaheuristics. As a well-established SOTA heuristic for VRPs, it serves as a reference standard to assess the solution quality and efficiency of our DAC-ALNS algorithm.
    
    \item AC-ALNS (Baseline): This variant represents a standard DRL paradigm (a single Actor-Critic framework coupled with ALNS) and serves as a baseline for our ablation study. By comparing it with our model, we can effectively decouple the effects of the dual mechanism, validating the need for a dual-actor architecture over the traditional single-actor approach.
\end{enumerate}

\subsubsection{Evaluation Metrics}

We employ three primary metrics to evaluate the solution quality and computational efficiency of the algorithm.

\begin{enumerate}
    \item Best Gap: This metric captures the algorithm's ultimate search potential, highlighting the performance gap between different methods based on their best-found solutions across independent runs.
    \item Avg Gap: The average gap across independent runs on the same instance is employed to evaluate how consistently the model performs, reflecting its overall solving stability.
    \item Time Ratio: This ratio quantifies the relative computational overhead, offering a direct efficiency benchmark against the baseline methods.
\end{enumerate}

To compare different algorithms, we measure the relative performance gap between a method we are testing (called $\mathcal{A}$) and a reference method (called $\mathcal{B}$). We calculate this gap like this:

\begin{equation}
\text{Gap}(\mathcal{A}, \mathcal{B}) = \frac{\text{Obj}_{\mathcal{A}} - \text{Obj}_{\mathcal{B}}}{\text{Obj}_{\mathcal{B}}} \times 100\%
\label{eq:gap_formula}
\end{equation}

Here, $\text{Obj}_{\mathcal{A}}$ and $\text{Obj}_{\mathcal{B}}$ are the best objective function values found by method $\mathcal{A}$ and method $\mathcal{B}$, respectively. A negative gap shows that method $\mathcal{A}$ is better than method $\mathcal{B}$ (it gives a better solution). If the Time Ratio is greater than 1, it means method $\mathcal{A}$ has a higher computational cost.

We use a one-sided Wilcoxon signed-rank test with a significance level of $\alpha=0.05$ to check if the performance differences are statistically significant.

\subsubsection{Implementation Details}

The proposed DAC-ALNS and all baseline algorithms were implemented in Python 3.10 and executed on a single CPU core. All experimental trials were conducted on a computing platform fitted with dual AMD EPYC 7Y43 processors (96 cores, operating at 2.25–3.35 GHz). Each problem instance was solved over 10 separate trials. The complete configuration of principal hyperparameters is summarized in Table~\ref{tab:hyperparameters}.

\begin{table}[!htbp]
\centering
\small 
\caption{Configuration of key hyperparameters}
\label{tab:hyperparameters}
\begin{tabular}{ll l} 
\toprule
\textbf{Category} & \textbf{Parameter} & \textbf{Value} \\
\midrule
\textbf{Training} 
 & Max. Iterations & $\max(2000, 50 \times n)$ \\
 & Random Seed Set & $\{0, 100, \dots, 900\}$ \\
 & Exploration Rate ($\epsilon$) & $0.1$ \\
\addlinespace[4pt] 
\textbf{ALNS Operators} 
 & Removal Ratio ($\rho$) & $[0.1, 0.4]$ \\
 & Worst Removal Count & $[5, 20]$ \\
 & Regret Repair ($N$) & $2$ \\
\addlinespace[4pt]
\textbf{DRL Agent} 
 & Learning Rate ($\alpha$) & $1 \times 10^{-3}$ \\
 & Discount Factor ($\gamma$) & $0.9$ \\
 & Entropy Coeff. & $0.01$ \\
\addlinespace[4pt]
\textbf{Reward \& State} 
 & Reward Terms $(r_1, r_2, r_3)$ & $(20, 10, -0.1)$ \\
 & GNN Embedding Dim. & $64$ \\
\bottomrule
\end{tabular}
\end{table}

\subsection{Comparative Experiments}

\subsubsection{Overall Performance Analysis}

Table~\ref{tab:comprehensive_results} presents the aggregate performance metrics across all CVRP and VRPTW instances, comparing DAC against ALNS and AC baselines.

\begin{table}[htbp]
  \centering
  \caption{Comprehensive performance comparison across CVRP and VRPTW instances}
  \label{tab:comprehensive_results}
  \setlength{\tabcolsep}{5pt}
  \begin{tabular}{llrrrrrrrrr}
    \toprule
    \multirow{2}{*}{Dataset} & \multirow{2}{*}{Size}
      & \multicolumn{3}{c}{DAC vs AC}
      & \multicolumn{3}{c}{DAC vs ALNS}
      & \multicolumn{3}{c}{AC vs ALNS} \\
    \cmidrule(lr){3-5}\cmidrule(lr){6-8}\cmidrule(lr){9-11}
      & & Best & Avg & Time & Best & Avg & Time & Best & Avg & Time \\
    \midrule
    \multirow{3}{*}{\textbf{CVRP}}
      & Small  & 0.00\%  & $-0.06\%$ & 1.05 & $-0.02\%$ & 0.01\%  & 2.33 & $-0.02\%$ & 0.07\%  & 2.28 \\
      & Medium & $-0.04\%$ & $-0.06\%$ & 1.32 & $-0.06\%$ & 0.01\%  & 2.23 & $-0.02\%$ & 0.07\%  & 1.69 \\
      & Large  & $-0.22\%$ & $-0.19\%$ & 1.58 & $-0.31\%$ & $-0.11\%$ & 2.47 & $-0.09\%$ & 0.07\%  & 1.57 \\
    \midrule
    \multirow{3}{*}{\textbf{VRPTW}}
      & Small  & $-0.01\%$ & 0.04\%  & 1.08 & 0.00\%  & $-0.04\%$ & 1.78 & 0.01\%  & $-0.08\%$ & 1.73 \\
      & Medium & $-0.10\%$ & $-0.57\%$ & 0.93 & $-0.11\%$ & $-0.49\%$ & 1.77 & $-0.01\%$ & 0.10\%  & 1.92 \\
      & Large  & $-0.03\%$ & $-0.08\%$ & 1.00 & $-0.14\%$ & $-0.44\%$ & 1.94 & $-0.11\%$ & $-0.37\%$ & 1.93 \\
    \bottomrule
  \end{tabular}
  \vspace{4pt}
  {\footnotesize\textit{Note:} Negative values indicate that the first-listed method outperforms the second. ${Time}(\mathcal{A}, \mathcal{B}) = \frac{T_{\mathcal{A}}}{T_{\mathcal{B}}}.$}
\end{table}

DAC continuously shows better solution quality than both baselines, as shown in Table~\ref{tab:comprehensive_results}. 
In the great majority of cases, DAC achieves negative Best and Avg Gaps in terms of solution quality. Interestingly, the overall trend shows that this advantage becomes more pronounced as the problem scale increases. For example, DAC performs $0.31\%$ (Best Gap) better than ALNS in Large CVRP. The improvement is even greater in VRPTW, where Avg Gaps reach $-0.44\%$ (Large) and $-0.49\%$ (Medium). In contrast to the suggested dual-actor framework, the AC baseline frequently displays positive Avg Gaps (e.g., $+0.10\%$ in Medium VRPTW), underscoring its instability.
Because DAC requires more policy inference and decision logic than AC and ALNS, it typically adds more overhead in terms of efficiency. 

 Nevertheless, this expense is not consistently higher in every situation. 
For example, on Medium VRPTW, DAC is faster than AC ($\text{TimeRatio}=0.93$), indicating that in some situations, the learned policy can minimize inefficient search steps. 
In general, the runtime results show a trade-off between search efficiency and decision overhead.
 However, DAC consistently produces better results under the same number of iterations, which at least implies that the choices it makes are more successful.

\subsubsection{Statistical Analysis of Comparative Performance}
Table~\ref{tab:statistical_summary} provides a detailed breakdown of the win/loss/tie records and statistical significance tests to verify the validity of these improvements and examine the performance trends.

\begin{table}[htbp]
  \centering
  \caption{Detailed statistical comparison: DAC vs. Baselines (ALNS \& AC)}
  \label{tab:statistical_summary}
  \small
  \setlength{\tabcolsep}{3.5pt}
  
  \begin{tabular}{llcccccccc}
    \toprule
    \multirow{2}{*}{\textbf{Dataset}} & \multirow{2}{*}{\textbf{Scale}} & \multirow{2}{*}{\textbf{$N$}} 
      & \multicolumn{3}{c}{\textbf{DAC vs. ALNS}} 
      & & \multicolumn{3}{c}{\textbf{DAC vs. AC (Ablation)}} \\
    \cmidrule(lr){4-6} \cmidrule(lr){8-10}
      & & & \textbf{W / L / T} & \textbf{$p$-val} & \textbf{Sig.} & & \textbf{W / L / T} & \textbf{$p$-val} & \textbf{Sig.} \\
    \midrule
    \multirow{4}{*}{\textbf{CVRP}} 
      & Small  & 18 & 5 / 6 / 7  & 0.5000 & n.s. & & 7 / 4 / 7  & 0.0653 & n.s. \\
      & Medium & 18 & 11 / 7 / 0 & 0.2475 & n.s. & & 13 / 5 / 0 & 0.1231 & n.s. \\
      & Large  & 18 & 12 / 6 / 0 & 0.0982 & n.s. & & 11 / 7 / 0 & 0.1419 & n.s. \\
      \cmidrule(lr){2-10}
      & \textbf{Global} & \textbf{54} & \textbf{28 / 19 / 7} & \textbf{0.0861} & \textbf{n.s.} & & \textbf{31 / 16 / 7} & \textbf{0.0298} & \textbf{*} \\
    \midrule
    \multirow{4}{*}{\textbf{VRPTW}} 
      & Small  & 18 & 6 / 7 / 5  & 0.5417 & n.s. & & 4 / 10 / 4 & 0.8016 & n.s. \\
      & Medium & 18 & 13 / 4 / 1 & 0.0123 & *    & & 12 / 4 / 2 & 0.0037 & ** \\
      & Large  & 18 & 15 / 3 / 0 & 0.0020 & **   & & 10 / 8 / 0 & 0.2899 & n.s. \\
      \cmidrule(lr){2-10}
      & \textbf{Global} & \textbf{54} & \textbf{34 / 14 / 6} & \textbf{0.0006} & \textbf{***} & & \textbf{26 / 22 / 6} & \textbf{0.0278} & \textbf{*} \\
    \bottomrule
    \multicolumn{10}{l}{\footnotesize \textit{Note:} \textbf{W/L/T}: Wins/Losses/Ties. \textbf{Sig.}: Significance based on Wilcoxon signed-rank test.} \\
    \multicolumn{10}{l}{\footnotesize Symbols: n.s. ($p > 0.05$); * ($p < 0.05$); ** ($p < 0.01$); *** ($p < 0.001$).} \\
  \end{tabular}
\end{table}

(1) DAC vs.\ ALNS (Absolute Performance).
The comparison suggests a clear scale effect. On small instances, DAC and ALNS are not statistically distinguishable (e.g., $p=0.5000$ on Small CVRP), and ties occur frequently. A likely reason is that these instances are relatively easy, so both methods often reach the best-known solutions and performance quickly saturates.
As the instances become larger and more constrained, DAC begins to separate itself from ALNS. On the Large VRPTW set, DAC wins 15 out of 18 runs ($p=0.0020$), which is unlikely to be explained by randomness alone. The aggregated VRPTW results show the same pattern: DAC significantly outperforms ALNS ($p=0.0006$). Overall, these findings indicate that DAC scales more effectively with increasing problem complexity.

(2) DAC vs.\ AC (Mechanism Effectiveness).
The ablation study against the simplified baseline (AC) highlights how instance complexity affects the usefulness of the proposed mechanism. On Small VRPTW, AC remains competitive and even wins 10 of 18 trials, indicating that a single-stream policy can be adequate when the underlying structure is relatively simple.
This advantage does not persist as the instances grow. On Medium and Large settings, DAC's dual-actor design becomes increasingly beneficial and the results shift in DAC's favor. For example, on Medium VRPTW DAC records 12 wins against 4 losses. When aggregated across benchmarks, DAC is significantly better than AC on both CVRP ($p=0.0298$) and VRPTW ($p=0.0278$), even though AC performs well on smaller cases. Overall, the evidence suggests that single-stream architectures may suffice at low complexity, whereas the proposed dual mechanism is important for coping with the larger, higher-dimensional search spaces in more challenging routing instances.

\subsection{Generalization Experiments}

To assess how well the proposed algorithm transfers to unseen problem instances, we carried out both \emph{in-distribution} and \emph{out-of-distribution} evaluations. We also consider a transfer variant, termed DAC-Transfer (DAC-T). For DAC-T, all network parameters are learned only on the source data and are then kept fixed during testing; that is, the model is used in a strictly inference-only setting on unseen instances.

For each source dataset, we split the data into a training portion ($2/3$) and a test portion ($1/3$). To reduce overfitting, we apply early stopping: training is halted if the best objective value does not improve for more than $2\%$ of the total iterations.

\subsubsection{Generalization Performance Results}
Table~\ref{tab:generalization_performance} details the performance gaps and time efficiency of the transfer model.

\begin{table}[htbp]
  \centering
  \caption{Generalization performance: DAC-T vs. Baselines}
  \label{tab:generalization_performance}
  \small
  \begin{tabular}{llcccccc}
    \toprule
    \multirow{2}{*}{\textbf{Exp. Type}} & \multirow{2}{*}{\textbf{Scenario}} & \multicolumn{3}{c}{\textbf{DAC-T vs. ALNS}} & \multicolumn{3}{c}{\textbf{DAC-T vs. DAC}} \\
    \cmidrule(lr){3-5} \cmidrule(lr){6-8}
     & & \textbf{Best Gap} & \textbf{Avg Gap} & \textbf{Time} & \textbf{Best Gap} & \textbf{Avg Gap} & \textbf{Time} \\
    \midrule
    \multirow{2}{*}{In-dist.} 
     & CVRP  & $-0.16\%$ & $-0.32\%$ & 2.23 & $-0.09\%$ & $-0.29\%$ & 0.99 \\
     & VRPTW & $0.00\%$  & $-0.07\%$ & 1.49 & $0.00\%$  & $-0.02\%$ & 0.85 \\
    \midrule
    \multirow{2}{*}{Out-of-dist.} 
     & CVRP  & $-0.60\%$ & $-0.57\%$ & 2.47 & $-0.09\%$ & $-0.52\%$ & 1.00 \\
     & VRPTW & $-0.38\%$ & $-0.72\%$ & 2.00 & $-0.05\%$ & $-0.29\%$ & 1.16 \\
    \bottomrule
    \multicolumn{8}{l}{\footnotesize \textit{Note:} Negative values indicate that DAC-T outperforms the baseline method. ${Time}(\mathcal{A}, \mathcal{B}) = \frac{T_{\mathcal{A}}}{T_{\mathcal{B}}}.$} \\
  \end{tabular}
\end{table}

A notable result in Table~\ref{tab:generalization_performance} is that DAC-T performs particularly well on out-of-distribution instances, in some cases even better than on in-distribution data. For example, on the out-of-distribution CVRP task, DAC-T improves upon ALNS by $-0.60\%$ in Best Gap. In addition, relative to the online-trained DAC, DAC-T delivers comparable (and consistently better, as indicated by negative gaps across all settings) solution quality while keeping the Time Ratio around, or below, 1.0. This suggests that DAC-T can achieve strong performance with excellent inference-time efficiency.

\subsubsection{Statistical Significance of Generalization}
To verify that these improvements are not due to random variation, we report statistical significance tests in Table~\ref{tab:generalization_stats}.

\begin{table}[htbp]
  \centering
  \caption{Statistical significance of generalization: DAC-T vs. Baselines}
  \label{tab:generalization_stats}
  \small
  \setlength{\tabcolsep}{3.5pt}
  
  \begin{tabular}{llcccc}
    \toprule
    \textbf{Comparison} & \textbf{Scenario} & \textbf{$N$} & \textbf{Win / Loss / Tie} & \textbf{$p$-val} & \textbf{Sig.} \\
    \midrule
    \multirow{4}{*}{DAC-T vs. ALNS} 
     & CVRP (In)   & 6 & 6 / 0 / 0 & 0.0156 & * \\
     & VRPTW (In)  & 6 & 4 / 1 / 1 & 0.1726 & n.s. \\
     & CVRP (Out)  & 8 & 7 / 1 / 0 & 0.0078 & ** \\
     & VRPTW (Out) & 8 & 8 / 0 / 0 & 0.0039 & ** \\
    \midrule
    \multirow{4}{*}{DAC-T vs. DAC} 
     & CVRP (In)   & 6 & 5 / 1 / 0 & 0.0312 & * \\
     & VRPTW (In)  & 6 & 2 / 2 / 2 & 0.3575 & n.s. \\
     & CVRP (Out)  & 8 & 7 / 1 / 0 & 0.0117 & * \\
     & VRPTW (Out) & 8 & 5 / 3 / 0 & 0.0547 & n.s. \\
    \bottomrule
    \multicolumn{6}{l}{\footnotesize \textit{Note:} \textbf{W/L/T}: Wins/Losses/Ties. \textbf{Sig.}: Significance based on Wilcoxon signed-rank test.} \\
    \multicolumn{6}{l}{\footnotesize Symbols: n.s. ($p > 0.05$); * ($p < 0.05$); ** ($p < 0.01$); *** ($p < 0.001$).} \\
  \end{tabular}
\end{table}

The statistical tests further support the stability of the transfer policy. On out-of-distribution instances, DAC-T wins all VRPTW cases against ALNS ($p=0.0039$) and wins 87.5\% of the CVRP cases ($p=0.0078$). More surprisingly, DAC-T is also significantly better than the online-trained DAC on CVRP (with $p=0.0312$ and $p=0.0117$), which suggests that the structural knowledge learned on small instances can be more useful than re-learning from scratch when the search space becomes high-dimensional.

\subsubsection{Transferability Insights and Mechanism Discussion}

DAC-T’s consistently strong results under distribution shift suggest that the policy is not merely memorizing size-dependent patterns, but instead learning operator-selection behavior that transfers across instance scales. A plausible explanation is that training on smaller instances helps the model acquire reusable structural priors for how ALNS operators should be chosen and combined. When the scale increases, online training typically becomes more difficult due to the higher-dimensional search space and sparser improvement signals, whereas DAC-T can exploit these learned priors directly at inference time. From an operational perspective, this supports a practical train small, infer large workflow, which is particularly appealing in time-sensitive logistics settings where extensive online learning is costly or simply impractical.

\section{Conclusion}

We introduce a DAC framework designed to improve both search efficiency and generalization when augmenting the classical ALNS metaheuristic for vehicle routing problems. Our key idea is to treat the destroy and repair phases as two coupled, but role-distinct, decision processes. With GNN-based state representations, the policy can exploit instance structure and adapt operator choices to the evolving solution trajectory, rather than relying solely on hand-crafted adaptive weight updates. Extensive experiments on standard benchmarks (CVRP and VRPTW) lead to the following conclusions:

\begin{enumerate}
    \item Strength in highly constrained settings: DAC remains competitive on CVRP and shows clear statistical advantages on VRPTW ($p < 0.001$). This indicates that the learned policy is particularly effective when the search landscape is shaped by tight time-window constraints. In these cases, optimizing long-horizon returns appears to encourage more forward-looking operator choices, reducing the short-sighted behavior often observed in conventional adaptive schemes.

    \item Value of the decoupled dual-actor design: The ablation study shows that the dual-actor structure consistently outperforms the single-stream baseline (AC). Separating destroy and repair allows each policy to specialize, which better matches the distinct combinatorial roles the two phases play within neighborhood search.

    \item Robust zero-shot transfer: A central finding is the transfer performance of DAC-T. Models trained only on small-scale instances and then used in a frozen, inference-only mode can scale to larger, out-of-distribution instances, and in several cases even exceed the performance of online, instance-specific training. This supports a practical ``\textit{train small, infer large}'' paradigm, reducing deployment-time computation and latency for real logistics use cases.
\end{enumerate}

Looking ahead, we plan to narrow the gap between benchmark assumptions and operational requirements. Building on the low-latency inference of DAC-T, we will extend the approach to richer VRP variants with realistic industrial constraints, including heterogeneous fleets, dynamic demand, and electric-vehicle routing considerations. In addition, integrating DAC into real-time decision-support pipelines will be important for evaluating robustness under the stochastic disruptions and uncertainty that characterize modern supply chains.

\bigskip

\textbf{Acknowledgements}

The research is supported by National Natural Science Foundation of China [grant no. 72301137]. We thank Cheng Hao for his efforts in the initial exploration.

\textbf{Data availability}

The repository https://github.com/Shaohua-Yu/dac-alns contains the data and source code for the methods that have been implemented.

\bibliographystyle{apalike}
\bibliography{review}

\appendix
\section{Detailed Action Space}
\label{A:dac}

The action space for CVRP and VRPTW employs a unified set of destroy and repair actions.

\textbf{(1) Destroy Actions}

\begin{itemize}
\item \textbf{Random Removal}: The algorithm randomly selects a certain proportion of nodes from the current solution based on a predefined removal ratio range and removes these nodes.
\item \textbf{Worst Removal}: The algorithm removes a certain proportion of nodes from the current solution in a greedy manner (removing nodes that make the worst contribution to the solution) based on a predefined removal ratio range.
\end{itemize}

\textbf{(2) Repair Actions}

\begin{itemize}
\item \textbf{Random Repair}: Random Repair randomly selects customers from the list of unserved customers and inserts them into any current feasible route, considering the paths of all vehicles during calculation.
\item \textbf{Greedy Repair}: Greedy Repair determines the cheapest insertion position for all currently unserved points, considering the paths of all vehicles during calculation.
\item \textbf{Regret Repair}: When reinserting removed nodes, this method finds the best current position and predicts future costs for a comprehensive optimization of the solution. This strategy has been proven to enhance the performance of the ALNS algorithm, especially in dealing with large-scale complex combinatorial optimization problems. Using the concept of “regret value,” the algorithm can explore the solution space more deeply, prevent premature convergence, and find a better global solution.
\end{itemize}

\end{document}